# Asking Better Questions - The Art and Science of Forecasting

A mechanism for truer answers to high-stakes questions

Emily Dardaman

BCG Henderson Institute, dardaman.emily@bcg.com

Abhishek Gupta

BCG Henderson Institute, gupta.abhishek@bcg.com

Without the ability to estimate and benchmark AI capability advancements, organizations are left to respond to each change reactively, impeding their ability to build viable mid and long-term strategies. This paper explores the recent growth of forecasting, a political science tool that uses explicit assumptions and quantitative estimation that leads to improved prediction accuracy. Done at the collective level, forecasting can identify and verify talent, enable leaders to build better models of AI advancements and improve inputs into design policy. Successful approaches to forecasting and case studies are examined, revealing a subclass of "superforecasters" who outperform 98% of the population and whose insights will be most reliable. Finally, techniques behind successful forecasting are outlined, including Phillip Tetlock's "Ten Commandments." To adapt to a quickly changing technology landscape, designers and policymakers should consider forecasting as a first line of defense.

## 1 INTRODUCTION

Across industry and the public sector, organizations are being caught off guard by advances in AI systems. Educators, public servants, and leaders of multinational organizations find themselves in the unpleasant position of needing to develop policies reactively to capability improvements, armed with neither a warning nor a glimpse of what technology can do next.

Organizations building advanced AI systems, like Microsoft, are not immune to significant surprises, as discovered during the Bing AI release [1]. AI has non-anthropomorphic traits that can make predictions difficult [2]. For example, new abilities or behaviors can be elicited from models well after their release.

Such surprises are suboptimal for everyone. To respond well to the increasing power of AI systems, organizations must develop forward-looking benchmarking systems and plan policy in advance. We must simultaneously create policies to govern the safe use of technologies with profound societal impacts, especially when integrated into society.

In this paper, we explore how a political science tool, forecasting, might enable leaders to build better models of AI development to use in crafting policy. Forecasting is making explicit, quantifiable predictions, using expressed assumptions and uncertainties to narrow in on the most likely outcomes. Transparency of reasoning allows others to challenge assumptions, provide additional information, and converge on the most likely outcomes.

German physicist Werner Heisenberg said, "We have to remember that what we observe is not nature [or AI] herself, but nature exposed to our method of questioning" [3]

If organizations can forecast AI advancements, they could make informed policy decisions about how to allocate capital, steward talent, and manage risks which will help them avoid unpleasant surprises.

## 1.1 SURPLUS OF UNEXAMINED EXPERTISE

Leaders are not operating in a void of expertise. Indeed, the advent of new technology creates a feeding frenzy for "expert insights" as surprised leaders attempt to gain their bearings. However, it can be difficult to sort genuine expertise from slick branding – and generally, culturally designated experts are not chosen exclusively for their prediction accuracy.

Instead, experts tend to be individuals optimized for three characteristics: reputation, experience, and communication skills. These are good qualities but are insufficient to determine whether individuals with high status will make the most accurate judgments on a given issue.

## 2. INTRODUCING FORECASTING

Forecasting is a decision-making tool that optimizes for accuracy. A forecast has three steps: defining a question with clear resolution criteria, submitting quantitative estimates, and evaluating the most correct estimate.

A classic example is election forecasting [4]. Elections are time-bound, uncertain, and officially adjudicated; forecasts can be right or wrong; individuals and organizations can compete over time to see who, with what methods, is most effective.

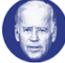

Figure 1: The Economist's 2020 US election forecast as of the final update on November 3, 2020. (https://projects.economist.com/us-2020-forecast/president)

**2.1 Improving forecasting**

The critical question is can we improve our prediction? The political costs of exchanging typical "experts" for leading forecasters are only worth it if forecasting works - if it is possible to improve judgment over time via processes that generalize to other areas.

Political scientist Phillip Tetlock, author of Superforecasting [21] and president of the Forecasting Research Institute, has dedicated his career to answering that question. In 2011, Tetlock and Barbara Mellers participated in a four-year experiment run by IARPA to predict difficult geopolitical and economic questions. Tetlock and Mellers recruited a team of online volunteers, competed against over 25,000 forecasters, and won handily [5], even against the intelligence agency's analysts.

They did this by identifying superforecasters [6]: the top 2% of forecasters from the pool of volunteers who consistently outperformed others and then putting these superforecasters in teams to collaborate with and sharpen each other.

Based on his interviews with these elite forecasters, Tetlock developed a model that leaders can learn from. Expertise is not the main predictor of success in forecasting; instead, it is the mindset and approach of the forecaster that separates good forecasters from bad ones. Superforecasters, Tetlock wrote, appreciate randomness and complexity and believe nothing is set in stone. They view beliefs as hypotheses to be tested, enjoy puzzles, and are quantitative thinkers. When facts change, they change their minds. Most importantly: they expect to grow.

**2.2 Forecasting markets**

Over time, forecasting allows better estimators to rise to the surface and for our best-shared thinking to map out possible outcomes. It can be used to identify talent, learn more about the problem, and gain strategic advantage.

Forecasting relies on quality and volume, so incentivizing people to participate is essential - either through compensation, which US gambling laws restrict [7], or reputation. Tetlock and his colleagues coined the term "superforecaster" to convey status and motivate top performers.

One way to achieve quality and volume is to create public prediction markets where forecasters worldwide can design questions, submit answers, and rank over time. These platforms can help leaders design policy for likely technical or geopolitical shifts.

Prediction markets, like Kalshi [8] and Polymarket [9], involve betting money on outcomes, where the bet's price functions as an aggregate measure of confidence in the outcome. Prediction aggregators, meanwhile, like Metaculus [10] and Good Judgement Open [11], use scoring methods to collect forecasts and assign reputation scores to lead forecasters. Some market makers, like Augur [12], operate globally through blockchain technology to avoid regional restrictions.

On public markets and aggregators, viewers can see 1) what the forecasters are claiming, 2) to what degree there is consensus on the issue, and 3) who has the best track record of performing well on similar questions.

Leaders can pay close attention to active markets with dedicated sections for AI outcomes, like Metaculus, which show live-updating confidence intervals on various AI topics from generative AI (deepfakes, AI art) to capability increases (computation benchmarks, time to human parity) and beyond.

Technology companies have already begun creating their markets. For the past year, Google has been running an internal prediction market with "over 175,000 predictions from over 10,000 Google employees [13]" The average success rate across 200 undisclosed but "difficult" questions rose and fell over time but closely tracked with actual events as they unfolded.

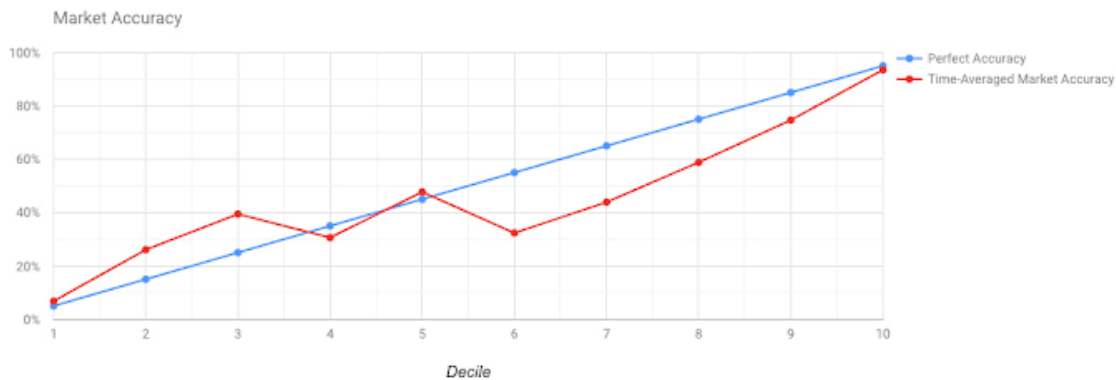

Figure 2: Forecasting accuracy of first 200 probability questions in Google's internal prediction market via Google [13]

Aggregating predictions this way can help achieve three goals:

1. Measure employee expectations and correct gaps in leaders' estimations
2. Surface superforecaster talent, who will be more likely to be accurate in the future
3. Once sufficiently confident in the forecasting talent, incorporate predictions into strategy.

Entering a forecast challenges leaders to clarify their position on a topic and provides accountability to the truth - both functions will help leaders iterate and improve their strategy. It allows leaders to test their underlying beliefs of what the world will look like.

## 3. FORECASTING AI

Forecasting can be about any topic, as long as it's clearly defined and resolvable and is promising for AI advances. On Metaculus, dozens of forecasters share and sharpen each other's explicit views on questions ranging from "Will the United States place restrictions on compute capacity before 2050?" [14] to "When will an AI achieve a 98th percentile score or higher in a Mensa admission test?" [15]  Meta-questions are also included, such as how likely forecasters will be surprised by AI progress [15], meaning they make a significant change to their forecasts en masse.

Intellectual humility is essential when forecasting AI. Research breakthroughs, availability of data and compute regulatory pressure, and geopolitical dynamics will all affect the speed at which AI develops. Forecasting offers not certainty but a chance to identify critical uncertainties and crowdsource the best available talent to whittle those down. In Tetlock's current work, he and his team are mapping AI uncertainties [16] into conditional trees – mapping the right questions to ask so that experts can begin disagreeing more productively.

Intelligence is the most powerful tool in the universe; advancing our understanding of its evolution will be the most important policy variable of our time.

### 3.1 Political headwinds

If forecasting has such significant benefits, why have policymakers not adopted it? The truth is that existing experts have little incentive to risk their positions by making falsifiable public claims. Many top performers in forecasting are relative unknowns without media presence or the weight of organizational support behind them.

Every form of accountability or transparency has faced similar hurdles in the past, but the significance and speed of AI capacity improvements give this old challenge new urgency. Our desire for accuracy must outweigh our affinity for popular pundits – who, again, are not categorically inaccurate in their predictions but have yet to be transparently tested.

### 3.2 How to forecast

In Superforecasting, Tetlock wrote, "Every consequential thing we do in the world is an implicit forecast because we're assuming what will follow later."

There is biological truth to that statement. Our brains model the world [17] through a constant process of prediction and preparation on both the cognitive and neural levels – it's one of the core functions of how we think. Projection allows us to stay grounded in the present, informed by the past, and align our behavior with the future we want to see. Wrong predictions have consequences not just for our beliefs but for our behavior.

In Superforecasting: The Art and Science of Prediction, and in his later work, Tetlock outlines how to improve our predictions. Good forecasters use a variety of strategies to improve their accuracy, such as breaking down complex problems into smaller parts, using probabilistic reasoning, seeking out diverse sources of information, and using feedback from mistakes to improve continuously. Coming off the success of the IARPA challenge, in 2015, superforecaster Warren Hatch opened Good Judgement, a company using forecasting to advise the World Economic Forum, the US Department of State Foreign Service Institute, investment banks, and risk management agencies. [18]

His complete list of ten commandments are:

> (1) Triage.
>
> (2) Break seemingly intractable problems into tractable sub-problems.

(3) Strike the right balance between inside and outside views.

(4) Strike the right balance between under and overreacting to evidence.

(5) Look for the clashing causal forces in each problem.

(6) Strive to distinguish as many degrees of doubt as the problem permits but no more.

(7) Strike the right balance between under- and overconfidence, prudence, and decisiveness.

(8) Look for the errors behind your mistakes but beware of rearview-mirror hindsight biases.

(9) Bring out the best in others and let others bring out the best in you.

(10) Master the error-balancing cycle.

(11) Don't treat commandments as commandments.

While these steps may look laborious, they allow policymakers to break intractable problems into easy parts using reference class analogies, attention to detail, and humility. Forecasting is about designing tighter feedback loops between technology design and our response to it, leading the way for proper technology-policy design integration.

**4. COLLECTIVE INTELLIGENCE**

The most potent problem-solving engine is not expertise or experience but collective intelligence (CI). We can define CI as what occurs when groups of individuals come together to "act intelligent:" to surface insights, solve problems, and converge on the truth. [19]

Wikipedia is an excellent example of CI at work, where thinkers worldwide can write, debate, and suggest edits at scale. Over time - despite attacks and misinformation [20] - it grows more comprehensive and accurate. The best insights are surfaced and protected. Like a famous adage in software programming called Linus's Law says, "Given enough eyeballs, all bugs are shallow."

When many people come together in a prediction market, they are bringing the power of CI to bear on a complex problem. Forecasting can help us understand how AI advances will inform policy and what kind of augmented CI we must look forward to, working in teams alongside advanced AI systems.

**5. CONCLUSION**

AI systems are becoming increasingly competent at everyday tasks, including complex and strategic work. These changes challenge policymakers across all sectors; changes will keep coming. Leaders face a dual challenge of improving their moment-to-moment access to decision-relevant information and designing policies that stand the test of time.

Forecasting is a powerful tool that can shorten the gap between technological development and policy response. Leaders should consider practicing the art of forecasting by breaking problems into their composite parts, making estimates, and charting their progress transparently across time. But even more importantly, they can contribute to publicly available prediction markets, elevate the voices of superforecasters, and ensure their industry's strategy is built on CI rather than shifting sands. Given that "superforecasters" are rare and that quantity improves accuracy, it is essential to encourage the development of public prediction markets and aggregators so that policymakers at all resource levels can incorporate predictions into their strategies.

**ACKNOWLEDGMENTS**

The BCG Henderson Institute funded and supported this work under Abhishek Gupta's Fellowship on Augmented Collective Intelligence.